# Upper critical field in a spin-charge separated superconductor.

R. G. Dias and J. M. Wheatley*

*Interdisciplinary Research Centre in Superconductivity,*
*University of Cambridge,*
*Madingley Road Cambridge, CB3 0HE, United Kingdom.*

(July 8, 1994)

It is demonstrated that the spatial decay of the pair propagator in a Luttinger liquid with spin charge separation contains a logarithmic correction relative to the free fermi gas result in a finite interval between the spin and charge thermal lengths. It is argued that similar effects can be expected in higher dimensional systems with spin charge separation and that the temperature dependence of the upper critical field $H_{c2}$ curve is a probe of this effect.

Recent experiments[1,2] on low transition temperature single layer cuprates indicate that the upper critical field curve in these materials has an anomalous upward curvature extending to temperatures below a small fraction of $T_c$. This behavior contrasts strongly with the mean field or weak coupling BCS result which gives an approximately parabolic shape for the $H_{c2}$ curve. This discrepancy is especially interesting because weak coupling theory could be expected to apply to systems with low $T_c$.[3]

Near to $T_c(H)$ the linearized gap equation becomes an homogeneous integral equation with a kernel which represents the normal state electron pair propagator in a magnetic field. Gorkov[4] included the effect of the gauge dependent phase acquired by a propagating electron, but assumed that the role of Landau quantization[5] could be neglected due to temperature or lifetime effects. Specializing to two dimensions, the linearized Gorkov equation in the symmetric gauge becomes:

$$\Delta(\mathbf{r}) = g \int d^2\mathbf{r}' K_0(\mathbf{r}', \beta) \exp(i\frac{\mathbf{r} \times \mathbf{r}'}{l^2}) \Delta(\mathbf{r} + \mathbf{r}') \quad (1)$$

where $\Delta(\mathbf{r})$ is the space dependent energy gap, $\beta$ is the inverse temperature, $g$ is the (assumed field independent) pairing interaction. The magnetic length $l$ is related to the applied field by $H = \phi_0(2\pi l^2)^{-1}$ where $\phi_0$ is the flux quantum $hc/e$. $K_0(\mathbf{r})$ is the fermion pair propagator or pair susceptibility in real space, in the absence of the external field and the pairing interaction $g$. Rajagopal and Vasudevan[6] pointed out the existence of a solution of Eq. 1 of the form $\Delta(\mathbf{r}) = \Delta_0 e^{-\frac{1}{2}(\frac{r}{l})^2}$. When substituted into Eq.1 this leads to the pairing instability condition

$$\frac{1}{2\pi g} = \int_{r_0}^{\infty} K_0(r, \beta) \exp\left(-\frac{r^2}{2l^2}\right) r dr \equiv \bar{K}_0(l, \beta) \quad (2)$$

where $r_0$ is a lower cutoff. $\bar{K}_0(l, \beta)$ is the uniform pair susceptibility in the presence of an external field.

Eq.2 is the mean-field instability condition defining the critical magnetic length $l_c(\beta)$. It states that *the upper critical field is a contour of constant pair susceptibility in the field-temperature plane.* The temperature dependence of $H_{c2}$ depends on the details of the normal state pair propagator. In the case of free fermions in D dimensions,

$$K_0^{\text{free}}(r, \beta) = \left(\frac{k_f}{2\pi r}\right)^{D-1} \frac{1}{v_f^2 \beta} \frac{1}{\sinh\left(\frac{2\pi r}{\beta v_f}\right)}. \quad (3)$$

When $r < \beta v_f = \xi$, the real space pair susceptibility decays as a power law $K_0^{\text{free}} \propto r^{-2}$ in 2D. At distances longer than the thermal length $\xi$ the pair propagator is exponentially small. Thus the integral Eq. 2 depends logarithmically on a long distance cutoff which is equal to $\xi$ or $l$, whichever is shorter. Substitution of Eq. 3 into Eq.2 leads to the conventional 2D almost parabolic $H_{c2}$ curve.

We can explore the consequences of an anomalous pair susceptibility via the ansatz $K_0(r, \beta) \sim r^{-\mu} \beta^{-\nu}$ for $r_0 < r < l$. For a 2D Fermi liquid we have seen that $\mu = 2$ and $\nu = 0$. The low temperature part of the $H_{c2}$ curve ($l \ll \xi$) is determined by this scaling. A weakly $r_0$ dependent result requires $\mu \leq 2$ and, with this assumption, Eq.2 gives $l_c \approx \beta^{\frac{\nu}{2-\mu}}$ or $H_{c2} \approx T^{\frac{2\nu}{2-\mu}}$. A power law "divergence" of $H_{c2}$ at low temperature occurs when $\nu < 0$.

The exponents $\nu, \mu$ can be directly related to scaling properties of the normal state single particle Green function $g(k, \omega)$ provided it is assumed that the vertex function has a trivial scaling. That is, we assume ($\omega_n = 2\pi n/\beta$),

$$K_0(r, \beta) = \beta^{-1} \sum_{\omega_n} g(r, \omega_n) g(r, -\omega_n). \quad (4)$$

Generalising an argument due to Balatsky[7], $g(\Lambda k, \Lambda \omega) = \Lambda^{-\alpha} g(k, \omega)$ implies that $K_0(r, \beta) = \beta^{-4-2\alpha} f(\frac{r}{\beta})$. Thus our scaling ansatz for $K_0(r, \beta)$ is recovered provided that $f(x) \sim x^{-\mu}$ which leads to

$$\nu = 4 + 2\alpha - \mu. \quad (5)$$

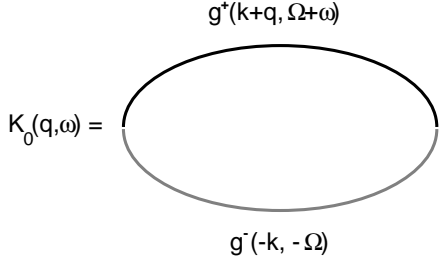

FIG. 1. The $q$-dependent pair susceptibility in a one dimensional Luttinger model with $g_4$ interaction only, whose Fourier transform is used in the text. Note that there are no vertex corrections in the absence of $g_2$ coupling between left and right branches.

Since $\mu \leq 2$, divergence of the upper critical field at low temperature is possible if $\alpha < -1$.[8]

While the experimental $H_{c2}$ curves show pronounced upward curvatures, the present data appear to be consistent with a finite upper critical field at zero temperature (i.e. $\nu = 0$). This suggests we pay special attention to the case of Fermi-liquid type scaling $\alpha = -1$. In addition to Fermi liquids, one-dimensional spinful Luttinger models with finite $g_4$ interaction only satisfy $\alpha = -1$. $g_4$ represents forward scattering of opposite spin electrons near the same fermi point. This produces spin charge separation but no anomalous scaling. For example, $Z$, the discontinuity in the fermion momentum distribution function at $k_f$, equals unity. It is natural to ask how the spatial dependence of the pair propagator is changed in this case. Note that microscopic models such as the 1D large $U$ Hubbard model give small but finite values of $g_2$, the forward scattering interaction between left and right movers.

The form of the 1D spin-charge separated retarded Green function in real space and time for right and left branches is[9,10]: $g^{\pm}(r,t) = -\frac{i}{2\pi}\Theta(t)\{[0^+ + i(x \mp v_s t)]^{-\frac{1}{2}}[0^+ + i(x \mp v_c t)]^{-\frac{1}{2}} + (x \to -x, t \to -t)\}e^{\pm ik_f r}$
where the spin and charge velocities are $v_{c,s} = v_f(1 \pm \frac{g_4}{2\pi})$. In momentum and frequency space, this is[10]

$$g^{\pm}(k,\omega) = \frac{1}{[i0^+ + \omega \mp v_s(k-k_f)]^{\frac{1}{2}}} \times \frac{1}{[i0^+ + \omega \mp v_c(k-k_f)]^{\frac{1}{2}}}. \quad (6)$$

The pair susceptibility (shown in Fig.1) can be expressed as,

$$K_0(r,\beta) = \frac{2}{\pi}\int d\omega \tanh(\beta\omega/2)\mathrm{Re}\{A(r,\omega)B(-r,-\omega)\} \quad (7)$$

where $A(r,\omega) = \frac{1}{2\pi}\int dk \exp(ikr)\mathrm{Im}g^+(k,\omega)$ and $B(r,\omega) = \frac{1}{2\pi}\int dk \exp(ikr)\mathrm{Re}g^+(k,\omega)$. Note that there is no vertex function in Eq.7 in the absence of $g_2$ coupling between left and right hand branches. We find that

$$A(r,\omega) = -\frac{1}{2\sqrt{v_s v_c}}e^{i\frac{1}{2}r\omega(v_s^{-1}+v_c^{-1})} \times$$
$$J_0\left(\frac{1}{2}r\omega(v_s^{-1} - v_c^{-1})\right)e^{ik_f r} \quad (8)$$

and $B(r,\omega) = i\mathrm{sgn}(r)A(r,\omega)$. For the free fermion case ($v_s = v_c$), $A(r,\omega)$ has an oscillatory, non-decaying behavior. When $v_s \neq v_c$ the Bessel function leads to an additional $r^{-\frac{1}{2}}$ decay beyond a distance $\approx v_s v_c (v_c - v_s)^{-1}\omega^{-1}$. This distance becomes large as the fermi surface is approached.

Eqs. 7 and 8 give

$$K(r,\beta) = \frac{1}{2\pi(v_c+v_s)}\frac{1}{r}\int_{-\infty}^{\infty} d\omega \tanh(\frac{\omega\bar{\xi}}{2r})\sin(\omega)J_0(\lambda\omega)^2 \quad (9)$$

where we have introduced the effective thermal length $\bar{\xi} = \beta\bar{v}$ where $\bar{v}^{-1} = v_s^{-1} + v_c^{-1}$ and the dimensionless parameter

$$\lambda = \frac{1}{2}\left|\frac{v_c - v_s}{v_c + v_s}\right|. \quad (10)$$

The $D = 1$ free fermion result Eq.3, is recovered upon setting $\lambda = 0$ and $\bar{v} = v_f/2$.[11]

The uniform pair susceptibility is obtained by integrating over $r$ with short distance cutoff $r_0 \ll \bar{\xi}$. Alternatively a high frequency cutoff can be introduced in Eq.9. Adopting the latter procedure, we find[12]

$$\bar{K}_0(T) = \frac{2}{\pi^2(v_c+v_s)}\boldsymbol{K}(2\lambda)\log[\frac{\beta\omega_c}{3.56}] \quad (11)$$

where $\boldsymbol{K}$ is the complete Elliptic function of the first kind. As $\lambda \to \frac{1}{2}$, $\boldsymbol{K}(2\lambda) \to -\frac{1}{2}\log(1-2\lambda) \approx \frac{1}{2}\log\left(\frac{v_c}{v_s}\right)$. Thus $g_4$ interactions *enhance* the pair susceptibility relative to the non-interacting result in this limit.

We have not succeeded in evaluating the pair propagator Eq.9 for general $\lambda$ in closed form. However progress can be made in the limit $v_s \ll v_c$, where departures from Fermi liquid behavior are most significant. In this limit the dominant contribution can be obtained using the substitution $\sin x J_0(\lambda x)^2 \to (\pi x)^{-1}\cos(1-2\lambda)x$ which is valid under the integral sign in Eq.9. Thus, in terms of spin ($\xi_s = \beta v_s$) and charge ($\xi_c = \beta v_c$) thermal lengths, we find

$$K_0(r,\beta) \simeq \frac{1}{\pi^2 v_c}\frac{1}{r}\begin{cases} \log\left[\frac{1}{\tanh[2\pi\frac{r}{\xi_c}]}\right] & r \gg \xi_s, \\ \log\left[\frac{v_c}{v_s}\right] & r \ll \xi_s. \end{cases} \quad (12)$$

The essence of this result is the opening up of a new regime in the window $\xi_s \ll r \ll \xi_c$. In this regime the pair susceptibility decays more slowly than the free fermion result $r^{-1}$ by a factor $\log(\xi_c/r)$. For $r \ll \xi_s, \xi_c$, the behavior is $r^{-1}$ with an enhancement factor $\log\left(\frac{v_c}{v_s}\right)$, while for $r \gg \beta v_s, \beta v_c$, $K_0(r,\beta)$ is exponentially small.

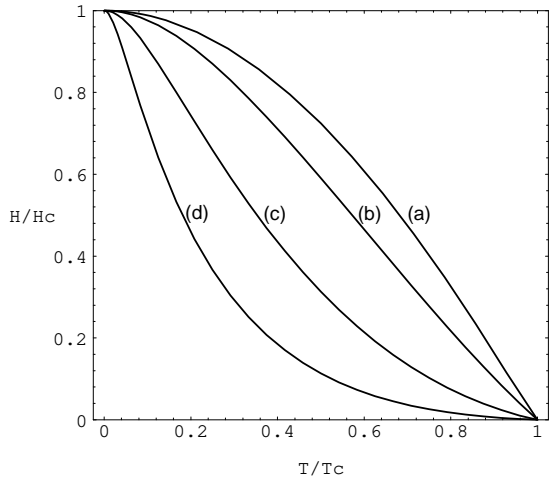

FIG. 2. Temperature dependence of the upper critical field on a normalized plot for (a) $v_s = v_c$, (b) $v_s = 0.1 v_c$, (c) $v_s = 0.01 v_c$, (d) $v_s = 0.001 v_c$.

We now ask whether an anomaly of this type, if present in a 2D system, would have an observable effect on $H_{c2}$. For example, if, following Anderson[14], we take the Green function Eq.6 as a model in $D = 2$, and continue to assume that vertex corrections can be neglected, then $K_0(r, \beta)$ has the form Eq.12 with the prefactor $r^{-1}$ replaced by $r^{-2}$.[15] We then solve Eq.2 for the critical field. The principle effects are: (1) The relationship between the zero temperature critical field and $T_c$ is altered compared to the Fermi liquid result. When $v_s = v_c$, this is $H_{c2}(0) = \frac{\pi}{\gamma} \phi_0 (T_c/v_f)^2$ where $\gamma = \exp(0.577)$.[13] When $v_s \ll v_c$ we find

$$H_{c2}(0) \approx \phi_0 \frac{T_c^2}{v_s v_c}. \qquad (13)$$

(2) $H_{c2}(0)$ is *enhanced* relative to the slope of the critical field curve near to $T_c$. When $v_s = v_c$ it is well known that $H_{c2}(0) \approx T_c \left|\frac{dH_{c2}}{dT}\right|_{T_c}$.[13] When $v_s \ll v_c$ we find

$$H_{c2}(0) \approx \frac{v_c}{v_s \log\left(\frac{v_c}{v_s}\right)} T_c \left|\frac{dH_{c2}}{dT}\right|_{T_c}. \qquad (14)$$

Numerical solution for dependence of $H_{c2}(T)$[16] is shown in Fig.2. As indicated above, spin-charge separation leads to upward curvature of the $H_{c2}$ line. Saturation behavior of the upper critical field always occurs at sufficiently low temperatures, when $l \ll \xi_c, \xi_s$. Conversely, linear variation of $H_{c2}$ with temperature occurs near zero field $l \gg \xi_c, \xi_s$.

In summary, the upper critical field is a probe of *both* spatial and temperature (frequency) dependence of normal state pair correlations. In a BCS superconductor the upper critical field satisfies the approximate relationship $a v_f^2 (H/\phi_0) + T^2 = T_c^2$ where $a$ is a numerical constant. This result is a direct manifestation of Fermi liquid theory for the normal state, which is characterized by a single quasiparticle velocity scale $v_f$. The effect of spin charge separation is visible in the upper critical field through the presence of distinct thermal lengthscales for spin and charge.

The authors wish to thank A. S. Alexandrov, A. P. McKenzie, J. R. Cooper and A. J. Schofield for important discussions. RD would like to thank Junta Nacional de Investigação Científica e Tecnológica (Lisbon) for financial support.

3